\documentstyle[11pt]{article}
\begin{document}

\title{\textbf{The generalized second law for the interacting generalized Chaplygin gas
model}}
\author{K. Karami$^{1,2}$\thanks{E-mail: KKarami@uok.ac.ir} ,
S. Ghaffari${^1}$, M.M. Soltanzadeh${^1}$\\$^{1}$\small{Department
of Physics, University of Kurdistan, Pasdaran St., Sanandaj,
Iran}\\$^{2}$\small{Research Institute for Astronomy
$\&$ Astrophysics of Maragha (RIAAM), Maragha, Iran}\\
}

\maketitle

\begin{abstract}
We investigate the validity of the generalized second law (GSL) of
gravitational thermodynamics in a non-flat FRW universe containing
the interacting generalized Chaplygin gas with the baryonic matter.
The dynamical apparent horizon is assumed to be the boundary of the
universe. We show that for the interacting generalized Chaplygin gas
as a unified candidate for dark matter (DM) and dark energy (DE),
the equation of state parameter can cross the phantom divide. We
also present that for the selected model under thermal equilibrium
with the Hawking radiation, the GSL is always satisfied throughout
the history of the universe for any spatial curvature, independently
of the equation of state of the interacting generalized Chaplygin
gas model.
\end{abstract}
\noindent{\textbf{Keywords:} dark energy theory . generalized Chaplygin gas}\\
%-----------------------------------------------------------------------------------------------
\clearpage
\section{Introduction}
One of interesting DE models is the Chaplygin gas (Kamenshchik et
al. 2000, 2001; Bilic et al. 2002) and its generalization (Bento et
al. 2002, 2004) which has been widely studied for interpreting the
accelerating universe. It is remarkable that the generalized
Chaplygin gas (GCG) equation of state has a well defined connection
with string theory and can be obtained from the light cone
parameterization of the Nambu-Goto action, associated with a D-brane
(Bilic et al. 2007). The GCG is the only gas known to admit a
supersymmetric generalization (Zhang et al. 2006). In this model, a
single self-interacting scalar field is responsible for both DE and
DM, giving also the observed accelerated expansion
(Garc\'{\i}a-Compe\'{a}n et al. 2008). The striking feature of the
GCG is that it allows for a unification of DE and DM. This point can
be easily seen from the fact that the GCG behaves as a dust-like
matter at early times and behaves like a cosmological constant at
late stage. This dual role is at the heart of the surprising
properties of the GCG model (Wu and Yu 2007; Lu et al. 2009).
Moreover, the GCG model has been successfully confronted with
various phenomenological tests involving SNe Ia data, CMB peak
locations, gravitational lensing and other observational data
(Alcaniz et al. 2003; Bento et al. 2003a,b; Bertolami et al. 2004;
Bento et al. 2005; Wu and Yu 2007; Lu et al. 2008; Lu et al. 2009).

Here our aim is to investigate the GSL of thermodynamics for the GCG
model as a candidate for the unified DM-DE which is in interaction
with the baryonic matter in the non-flat universe enclosed by the
apparent horizon.
%-----------------------------------------------------------------------------------------------
\section{Interacting GCG and BM}
The GCG model is based on the equation of state
\begin{equation}
P_{\rm Ch}=-\frac{A}{\rho_{\rm Ch}^{\alpha}}, \label{GCG}
\end{equation}
where $\rm A$ and $\alpha$ are the GCG constant parameters (Bento et
al. 2002, 2004). The case $\alpha=1$ corresponds to the standard
Chaplygin gas model (Kamenshchik et al. 2000, 2001; Bilic et al.
2002).

Using Eq. (\ref{GCG}), the continuity equation can be integrated to
give
\begin{equation}
\rho_{\rm Ch}=\rho_{\rm
Ch0}\left[A_{s}+\frac{(1-A_{s})}{a^{3(1+\alpha)}}\right]^{\frac{1}{1+\alpha}},
\label{eqroh}
\end{equation}
where $A_{s}\equiv\frac{A}{\rho_{\rm Ch0}^{1+\alpha}}$, $\rho_{\rm
Ch0}$ is the present energy density of the GCG, and $a$ is the
cosmic scale factor. Note that the GCG model smoothly interpolates
between a non-relativistic matter phase ($\rho_{\rm Ch}\propto
a^{-3}$) in the past and a negative-pressure DE regime ($\rho_{\rm
Ch}=-P_{\rm Ch}$) at late times. This interesting feature leads to
the GCG model being proposed as a candidate for the unified DM-DE
(UDME) scenario (Wu and Yu 2007; Lu et al. 2009).

In the framework of the Friedmann-Robertson-Walker (FRW) metric
\begin{equation}
{\rm d}s^2=-{\rm d}t^2+a^2(t)\left(\frac{{\rm d}r^2}{1-kr^2}+r^2{\rm
d}\Omega^2\right),\label{metric}
\end{equation}
the first Friedmann equation has the following form
\begin{equation}
{\textsl{H}}^2+\frac{k}{a^2}=\frac{8\pi}{3}~ (\rho_{\rm
Ch}+\rho_{\rm b}),\label{eqfr}
\end{equation}
where we take $G=1$ and $k=0,1,-1$ represent a flat, closed and open
FRW universe, respectively. Also $\rho_{\rm Ch}$ and $\rho_{\rm b}$
are the energy density of GCG and BM, respectively. In terms of the
the dimensionless energy densities as
\begin{equation}
\Omega_{\rm b}=\frac{8\pi\rho_{\rm b}}{3H^2},~~~\Omega_{\rm
Ch}=\frac{8\pi\rho_{\rm Ch}}{3H^2},~~~\Omega_{k}=\frac{k}{a^2H^2},
\label{eqomega}
\end{equation}
the first Friedmann equation yields
\begin{equation}
\Omega_{\rm b}+\Omega_{\rm Ch}=1+\Omega_{k}.\label{eq10}
\end{equation}
The energy equations for the GCG and BM with $\omega_{\rm b}=0$ are
\begin{equation}
\dot{\rho}_{\rm Ch}+3H(1+\omega_{\rm Ch})\rho_{\rm
Ch}=-Q,\label{eqpol}
\end{equation}
\begin{equation}
\dot{\rho}_{\rm b}+3H\rho_{\rm b}=Q,\label{eqCDM}
\end{equation}
where following Kim et al. (2006), we choose $Q=3b^2H(\rho_{\rm
Ch}+\rho_{\rm b})$. The interaction between the GCG and BM can be
detected by local gravity measurements (Hagiwara et al. 2002;
Peebles and Ratra 2003). Following Guendelman and Kaganovich (2008),
the interaction between the BM and the DE will cause the appearance
of the fifth force. However, according to the results of the fifth
force experiments, a coupling of the DE (modeled by a light scalar
field) to visible (baryonic) matter is strongly suppressed.
According to Guendelman and Kaganovich (2008), a new aspect
introduced by modern cosmology to this problem is the question of
why the coupling of the light scalar (DE) to visible matter is
strongly suppressed while similar coupling to DM is energetic.
Discovery of DE and cosmic coincidence interpreted as evidence of
the existence of an unsuppressed DE to DM coupling, turns the fifth
force problem into an actual and even burning fundamental puzzle.

Taking time derivative of Eq. (\ref{eqroh}) yields
\begin{equation}
\dot{\rho}_{\rm Ch}=-3(1-A_s)H\frac{\rho_{\rm
Ch0}}{a^{3(1+\alpha)}}\left[A_{s}+\frac{(1-A_{s})}{a^{3(1+\alpha)}}\right]^{\frac{-\alpha}{1+\alpha}}.\label{teqroh}
\end{equation}
Substituting Eq. (\ref{teqroh}) in (\ref{eqpol}) gives the equation
of state (EoS) parameter of the interacting GCG model as
\begin{equation}
\omega_{\rm
Ch}=-\frac{A_s{a^{3(1+\alpha)}}}{1-A_{s}+A_{s}a^{3(1+\alpha)}}-b^2\Big(\frac{1+\Omega_{k}}{\Omega_{\rm
Ch}}\Big)\label{omegach}.
\end{equation}
Observational constraints on the GCG model as the UDME from the
joint analysis of the latest astronomical data give the best-fit
values of the GCG model parameters. In this respect, Wu and Yu
(2007) obtained $(A_{\rm s}=0.76, \alpha=-0.005)$ for non-flat
universe with $(\Omega_k=0.04, \Omega_{\rm Ch}=0.995)$. Also Lu et
al. (2009) derived $(A_{\rm s}=0.73, \alpha=-0.09)$ for flat
universe with $(\Omega_{\rm Ch}=0.956)$. Equation (\ref{omegach})
shows that in the absence of interaction between GCG and BM, $b^2 =
0$, and for the present time, $a=1$, $\omega_{\rm Ch}=-A_s>-1$ and
cannot cross the phantom divide. However, in the presence of
interaction, $b^2\neq 0$, taking $(A_{\rm s}=0.76, \alpha=-0.005,
\Omega_k=0.04, \Omega_{\rm Ch}=0.995)$ given by Wu and Yu (2007),
$(A_{\rm s}=0.73, \alpha=-0.09, \Omega_k=0.0, \Omega_{\rm
Ch}=0.956)$ given by Lu et al. (2009) and $a=1$ for the present
time, Eq. (\ref{omegach}) clears that the phantom EoS, i.e.
$\omega_{\rm Ch}<-1$, can be achieved when $b^2>0.23$ and $b^2>0.26$
for non-flat and flat universe, respectively.

The deceleration parameter is given by
\begin{equation}
q=-\Big(1+\frac{\dot{H}}{H^2}\Big).\label{q1}
\end{equation}
Taking time derivative of both sides of Eq. (\ref{eqfr}), and using
Eqs. (\ref{eqomega}), (\ref{eq10}), (\ref{eqpol}) and (\ref{eqCDM}),
we get
\begin{equation}
q=\frac{1}{2}\Big(1+\Omega_{k}+3\Omega_{\rm Ch}\omega_{\rm
Ch}\Big).\label{qdec2}
\end{equation}
Substituting Eq. (\ref{omegach}) in (\ref{qdec2}) yields
\begin{equation}
q=\frac{1}{2}\Big[\frac{-3A_{s}a^{3(1+\alpha)}}{1-A_{s}+A_{s}a^{3(1+\alpha)}}\Omega_{\rm
ch}+(1-3b^2)(1+\Omega_{k})\Big].\label{q3}
\end{equation}
Now taking $(A_{\rm s}=0.76, \alpha=-0.005, \Omega_k=0.04,
\Omega_{\rm Ch}=0.995)$ given by Wu and Yu (2007), $(A_{\rm s}=0.73,
\alpha=-0.09, \Omega_k=0.0, \Omega_{\rm Ch}=0.956)$ given by Lu et
al. (2009) and $a=1$ for the present time we get $q=-0.61-1.56b^2$
and $q=-0.55-1.5b^2$ for non-flat and flat universe, respectively.
These results show that the deceleration parameter is always
negative even in the absence of interaction between GCG and BM.
Therefore the GCG model in the present time can drive the universe
to accelerated expansion.
%-----------------------------------------------------------------------------------------------
\section{GSL of thermodynamics}
Here, we study the validity of the GSL in which the entropy of the
GCG and BM inside the horizon plus the entropy of the horizon do not
decrease with time (Wang et al. 2006).

The location of the apparent horizon $\tilde{r}_{\rm A}$ in the FRW
universe according to Cai et al. (2009) is obtained as
\begin{equation}
\tilde{r}_{\rm A}=H^{-1}(1+\Omega_{k})^{-1/2}.\label{ah}
 \end{equation}
For $k = 0$, the apparent horizon is same as the Hubble horizon,
i.e. $\tilde{r}_{\rm A}=H^{-1}$.

Following Cai and Kim (2005), the associated Hawking temperature on
the apparent horizon is given by
\begin{equation}
T_{\rm A}=\frac{1}{2\pi \tilde{r}_{\rm
A}}\Big(1-\frac{\dot{\tilde{r}}_{\rm A}}{2H\tilde{r}_{\rm A}}
\Big),\label{TA1}
\end{equation}
where $\frac{\dot{\tilde{r}}_{A}}{2H\tilde{r}_{A}}<1$ ensure that
the temperature is positive.

Taking time derivative of both sides of (\ref{ah}) and using Eqs.
(\ref{eqfr}), (\ref{eqomega}), (\ref{eq10}), (\ref{eqpol}),
(\ref{eqCDM}), (\ref{q1}) and (\ref{qdec2}), one can rewrite Eq.
(\ref{TA1}) in real dimension form as
\begin{eqnarray}
T_{\rm A}&=&\frac{\hbar c}{k_{\rm
B}}\frac{H}{8\pi(1+\Omega_k)^{1/2}}(1+\Omega_{k}-3\Omega_{\rm
Ch}\omega_{\rm Ch}),\nonumber\\
&=&\frac{\hbar c}{k_{\rm
B}}\frac{H}{4\pi(1+\Omega_k)^{1/2}}(1+\Omega_{k}-q),\label{TA2}
\end{eqnarray}
where $\hbar$, $c$ and $k_{\rm B}$ are the reduced Planck constant,
speed of light and Boltzmann constant, respectively. Now taking
$H_0=72~{\rm Km~s^{-1} Mpc^{-1}}$ for the present time, one can
estimate the Hawking temperature on the apparent horizon as $T_{\rm
A_0}\sim\frac{\hbar cH_0}{k_{\rm B}}\sim 10^{-22}~{\rm K}$.

Following Maartens (1996), the BM (non-–relativistic matter)
temperature in the universe scales as $T\propto a^{-2}$, so like
Gong et al. (2007b) we assume here that the GCG temperature has a
similar behavior $T\propto a^{-n}$ to avoid the negative entropy
problem, where $n$ is an arbitrary constant. It is not necessary to
take $n=2$ to ensure that the GCG is in equilibrium with the BM,
since their dispersion relations could be completely different (see
Lima and Alcaniz 2004; Gong et al. 2007a). Since the usual BM
temperature in the universe decreases as the universe expands, we
expect that the GCG temperature also preserves this property. Santos
et al. (2006) showed that the GCG temperature in the adiabatic
evolution from a dust-like to a de Sitter cosmological model remains
in the range $0<T <T_*$, where $T_*\sim 10^{32}~{\rm K}$, the
temperature of the Planck era, is the maximum temperature of the GCG
when it fills small volumes. Therefore the temperature of the
universe filled with the GCG cools down as expected. This shows that
the temperature behaviour of the GCG is different from that of the
DE with constant EoS parameter $\omega$. The temperature of the GCG
decreases instead of increasing. This tells us that the increasing
or decreasing behaviour of the temperature of the universe dominated
by DE is model dependent; it is not a general property associated
with the DE (Gong et al. 2007a). A basic difficulty, however, is
that the present-day DE temperature has not been measured. Lima and
Alcaniz (2004) using a very naive estimate obtained the present
value of the DE temperature as $T_{\rm DE}^0\sim 10^{-6}~{\rm K}$ in
a non-flat FRW universe and in the absence of interaction with BM.
Zhou et al. (2009) showed that the DM temperature in the absence of
interaction with the DE behaves same as the BM as $T_{\rm DM}\propto
a^{-2}$. They estimated $T_{\rm DM}^0\sim 10^{-7}~\rm K$ for the
present time. Since the temperatures of the DE and the BM at the
present time differs from that of the horizon $T_{\rm A_0}\sim
10^{-22}~{\rm K}$, the systems must interact for some length of time
before they can attain thermal equilibrium. Although in this case
the local equilibrium hypothesis may no longer hold (Das et al.
2002, Wang et al. 2008; Pav\'{o}n and Wang 2009; Zhou et al. 2009),
Karami and Ghaffari (2010) showed that the contribution of the heat
flow between the horizon and the fluid in the GSL in non-equilibrium
thermodynamics is very small, $O(10^{-7})$. Therefore the
equilibrium thermodynamics is still preserved.

The entropy of the universe containing the GCG and BM is given by
Gibb's equation (Izquierdo and Pav\'{o}n 2006a)
\begin{equation}
T{\rm d}S={\rm d}E+P{\rm d}V,\label{eqSLT1}
\end{equation}
where ${\rm V}=4\pi \tilde{r}_{\rm A}^3/3$ is the volume of the
universe and $T=T_{\rm A}$. Also
\begin{equation}
E=\frac{4\pi \tilde{r}_{\rm A}^3}{3} (\rho_{\rm Ch}+\rho_{\rm
b}),\label{eqEde}
\end{equation}
\begin{equation}
P=P_{\rm Ch}+P_{\rm b}=P_{\rm Ch}=\omega_{\rm Ch}\rho_{\rm
Ch}=\frac{3H^2}{8\pi}\Omega_{\rm Ch}\omega_{\rm Ch}. \label{eqEcdm}
\end{equation}
Taking time derivative of both sides of (\ref{eqSLT1}) and using
Eqs. (\ref{eqfr}), (\ref{eqomega}), (\ref{eq10}), (\ref{eqpol}),
(\ref{eqCDM}), (\ref{eqEde}) and (\ref{eqEcdm}), we obtain
\begin{eqnarray}
T_{\rm A}\dot{S}=-4\pi H\tilde{r}_{\rm A}^3\rho_{\rm
Ch}(1+u+\omega_{\rm Ch})\Big(1-\frac{\dot{\tilde{r}}_{\rm
A}}{H\tilde{r}_{\rm A}}\Big),\label{Sah}
\end{eqnarray}
where $u=\rho_{\rm b}/\rho_{\rm Ch}$.

Also the evolution of the geometric entropy on the apparent horizon
$S_{{\rm A}}=\pi \tilde{r}_{\rm A}^{2}$ (Izquierdo and Pav\'{o}n
2006a) is obtained as
\begin{eqnarray}
T_{\rm A}\dot{S}_{\rm A}=4\pi H\tilde{r}_{\rm A}^3\rho_{\rm
Ch}(1+u+\omega_{\rm Ch})\Big(1-\frac{\dot{\tilde{r}}_{\rm
A}}{2H\tilde{r}_{\rm A}}\Big).\label{SAah}
\end{eqnarray}
Finally, adding Eqs. (\ref{Sah}) and (\ref{SAah}) yields the GSL as
\begin{equation}
T_{\rm A}\dot{S}_{\rm tot}=2\pi\tilde{r}_{\rm A}^2\rho_{\rm
Ch}(1+u+\omega_{\rm Ch})\dot{\tilde{r}}_{\rm A},\label{Stotah}
\end{equation}
where $S_{\rm tot}=S+S_{\rm A}$ is the total entropy.

Taking time derivative of the apparent horizon (\ref{ah}) and using
Eq. (\ref{eqfr}), one can obtain
\begin{eqnarray}
\dot{\tilde{r}}_{\rm A}=4\pi H\tilde{r}_{\rm A}^3\rho_{\rm
Ch}(1+u+\omega_{\rm Ch})\label{ahdot}.
\end{eqnarray}
Substituting Eq. (\ref{ahdot}) into Eq. (\ref{Stotah}) gives
\begin{equation}
T_{\rm A}\dot{S}_{\rm tot}=8\pi^2H\tilde{r}_{\rm A}^5\rho_{\rm
Ch}^2(1+u+\omega_{\rm Ch})^2\geq 0,\label{Stotah1}
\end{equation}
which shows that the GSL for the universe containing the interacting
GCG with BM enclosed by the dynamical apparent horizon is always
satisfied throughout the history of the universe for any spatial
curvature, independently of the EoS parameter of the interacting GCG
model.
%-----------------------------------------------------------------------------------------------
\\
\\
\noindent{{\bf Acknowledgements}}\\ The authors thank the unknown
referee for very valuable comments. The work of K. Karami has been
supported financially by Research Institute for Astronomy $\&$
Astrophysics of Maragha (RIAAM), Maragha, Iran.
%-----------------------------------------------------------------------------------------------

%-----------------------------------------------------------------------------------------------

\end{document}